\documentclass{PoS}
\newcommand{\lsi}{LS~I~+61 303 }

\newcommand{\uu}{4U 1036--56 }
\newcommand{\f}{1FGL J1018.6-5856 }
\newcommand{\is}{IBIS/ISGRI }
\newcommand{\je}{JEM--X }
\newcommand{\h}{HESS J0632+057}
\newcommand{\agl}{AGL J2241+4454 }

\title{\emph{INTEGRAL} $\&$ \emph{RXTE} View of Gamma-ray Binaries }

\ShortTitle{INTEGRAL $\&$ RXTE View of Gamma-ray Binaries}

\author{\speaker{Jian Li}\\
        Institute of High Energy Physics, Chinese Academy of Sciences, China\\
        Institut de Ci\`encies de l'Espai (IEEC-CSIC), Spain\\
        E-mail: \email{jianli@ihep.ac.cn}}

\author{Diego F. Torres\\
        Institut de Ci\`encies de l'Espai (IEEC-CSIC), Spain\\
              Instituci\'o Catalana de Recerca i Estudis Avan\c{c}ats (ICREA)\\
        E-mail: \email{dtorres@ieec.uab.es}}

\author{Shu Zhang\\
        Institute of High Energy Physics, Chinese Academy of Sciences, China\\
        E-mail: \email{szhang@ihep.ac.cn}}

\author{Jianmin Wang\\
        Institute of High Energy Physics, Chinese Academy of Sciences, China\\
        National Astronomical Observatories of China, Chinese Academy of Sciences, China\\
        E-mail: \email{wangjm@ihep.ac.cn}}

\abstract{Gamma-ray binaries are X-ray binaries with gamma-ray emissions. Their multi-wavelength emissions range from radio, optical, X-ray and to very high energy (TeV). X-ray emissions are crucial to understand the nature of gamma-ray binaries. \emph{INTEGRAL} and \emph{RXTE} have covered and monitored most of the gamma-ray binaries in hard and soft X-rays. Here we report the results of several gamma-ray binaries and possible gamma-ray binaries from \emph{INTEGRAL} and \emph{RXTE}.}

\FullConference{"An INTEGRAL view of the high-energy sky (the first 10 years)"
9th INTEGRAL Workshop and celebration of the 10th anniversary of the launch,\\
		October 15-19, 2012\\
		Bibliotheque Nationale de France, Paris, France}

\begin{document}
\section{Introduction }

Various classes of gamma-ray sources have been detected up to the TeV band. Among them, the class of gamma-ray emitting binaries
 focuses great attention. They are X-ray binaries hosting O/B companions, which have gamma-ray emission up to High-Energy
 (HE, E$\textgreater$100 MeV) and/or Very High-Energy (VHE, E$\textgreater$100 GeV), modulated on the orbital period.
Only a handful of such binaries are known (e.g., LS 5039, LSI +61 303, PSR B1259-63, HESS J0632+057, 1FGL J1018.6-5856, or Cyg X-3), although a larger population is expected. To understand their nature and particle acceleration mechanisms, X-ray emissions of gamma-ray binaries are crucial. \emph{INTEGRAL} and \emph{RXTE} have observed and monitored most of the gamma-ray binaries in hard and soft X-ray. Here we present the results for several confirmed (\lsi, \f and \h ) and possible gamma-ray binaries (AGL J1037--5708 and \agl) in the view of \emph{INTEGRAL} and \emph{RXTE}.

\section{Super-orbital modulation of \lsi }

Our data set includes 473 \emph{RXTE/PCA} pointed observations from 2007 August 28 to 2011 September 15.
The analysis is performed using the standard \emph{RXTE/PCA} criteria. Only PCU2 has been used for the analysis. Our
 count rate values are given for an energy range of 3-30 keV. In order to remove the influence of several kilosecond-long flares, we cut all observations that presented a larger count rate than three times the average.

Given a six-month time bin, we take the peak X-ray flux in orbital lightcurve and
compute the modulated flux fraction. The latter is defined as
$(c_{max} - c_{min})/(c_{max} + c_{min})$, where $c_{max}$ and $c_{min}$
are the maximum and minimum count rates in the 3-30 keV orbital lightcurve of
that period. Results are shown in Figure 1. Table 1 presents
the values of the reduced $\chi^{2}$ for fitting different models to the
modulation fraction and the peak flux in X-rays. It compares the
results of fitting a horizontal line, a linear fit, and two sinusoidal
functions. One of the latter has the same period and phase of
the radio modulation (from Gregory 2002, labeled as ``Radio" in Table 1, dotted line in Figure 1). The other sine function has
the same period as in radio but allowing for a phase shift from it
(a solid line in Figure 1, labeled as ``Shifted" in Table 1). It is
clear that there is variability in the data and the sinusoidal description with a phase
shift is better than the linear one. The phase shift derived by fitting the modulated
fraction is 281.8$\pm$44.6 days, corresponding in phase
to $\sim$0.2 of the 1667$\pm$8 day super-orbital period. The phase
shift derived by fitting the maximum flux is 300.1$\pm$39.1 days,
which are compatible with the former.

This super--orbital modulation is also
reported in Chernyakova et al. 2012.
We show that there is a $\sim0.2$ phase shift between the radio and the X-ray super-orbital modulation. Torres et al. 2012 has recently discussed that \lsi\ could be subject to
a flip-flop behavior. The super-orbital modulation is possibly due to
the cyclic change of the circumstellar disk (Li et al. 2012a; see also
Papitto et al. 2012). In this context, multi--wavelength super--orbital modulation is expected and confirmed in radio, optical, X-ray and hinted in GeV and TeV (Li et al. 2012a; Hadasch et al. 2012).

\begin{figure*}[t]
\centering
  \includegraphics[angle=0, scale=0.375] {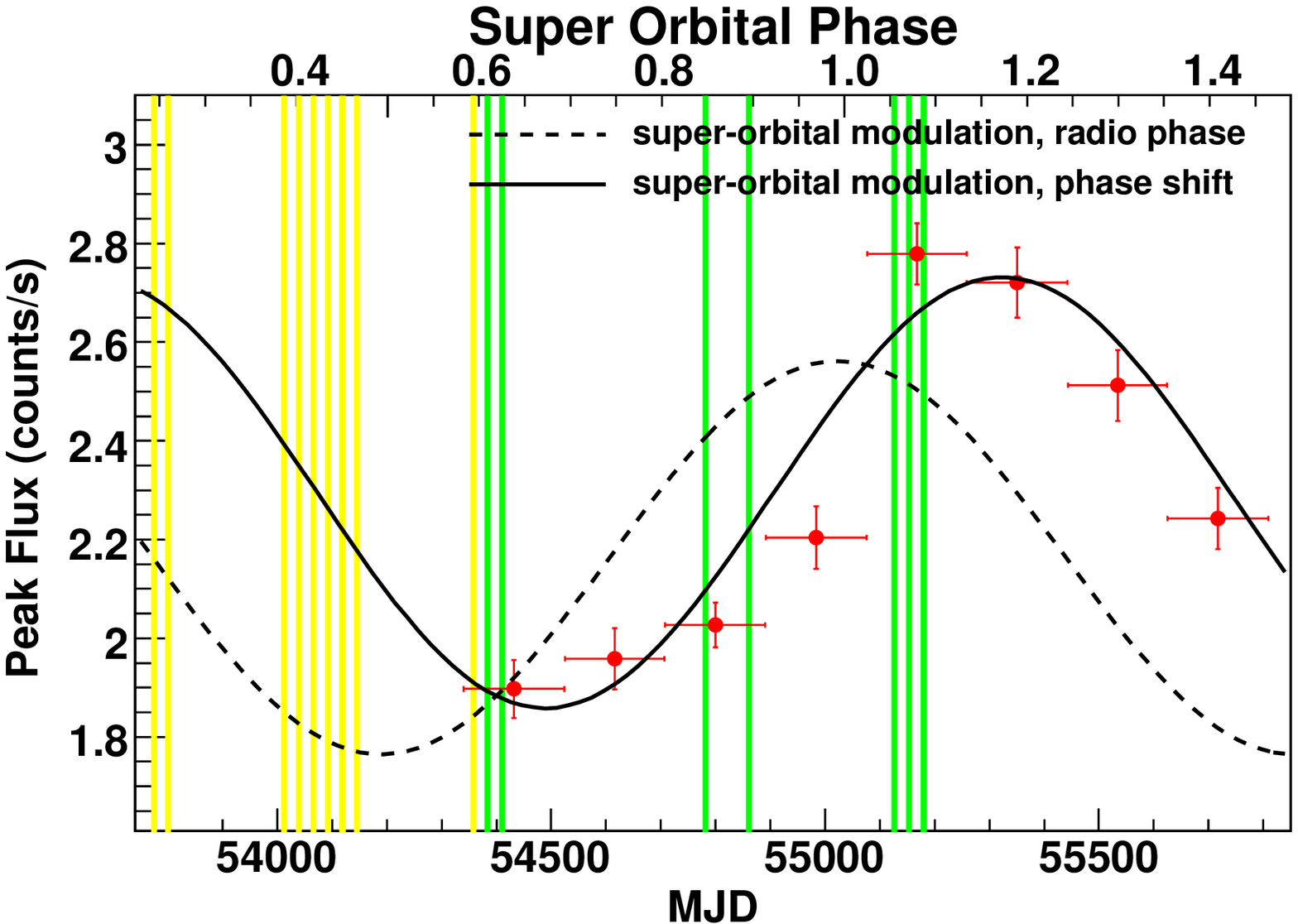}
  \includegraphics[angle=0, scale=0.375] {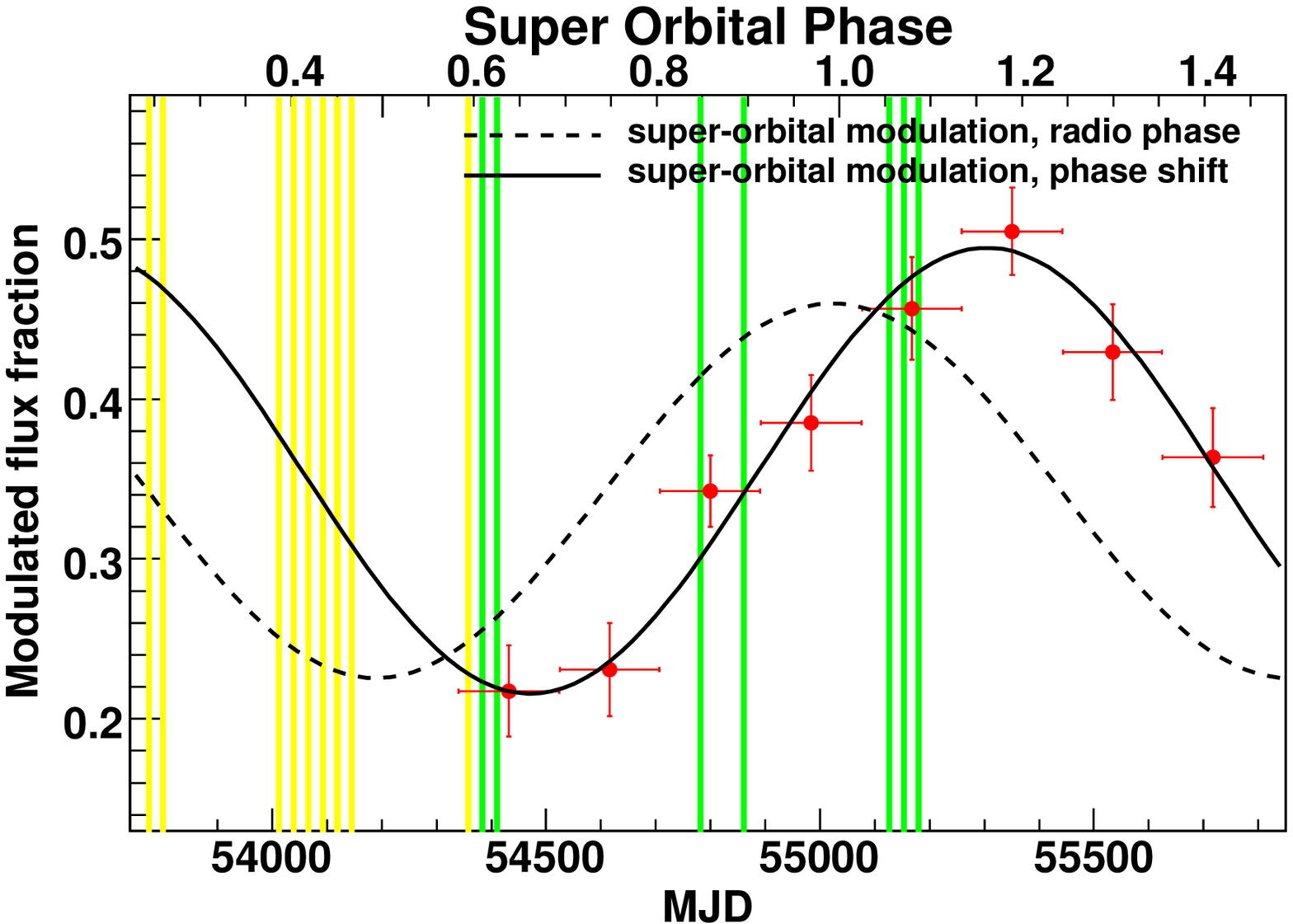}
 \caption{Left: Peak count rate of the X-ray emission from \lsi\ as a function of time and the super-orbital phase.
 Right: modulated fraction, see text for details. The dotted line shows the sine fitting to the modulated flux fraction and peak flux with a period and phase fixed at the radio parameters (from Gregory 2002).
The solid curve stands for sinusoidal fit obtained by fixing the period at the 1667 days value, but letting the phase vary. The time bin corresponds to six months. The colored boxes represent the times of the TeV observations that covered the broadly-defined apastron region. The boxes in green denote the times when TeV observations are in low state while boxes in yellow are TeV observations in high state.
}
\label{super-orb}
\end{figure*}

\section{INTEGRAL detection of \f}

\f is a newly discovered gamma-ray binary by \emph{Fermi/LAT}. It has an orbital period of $16.58 \pm
0.04$ days and an epoch of maximum gamma-ray flux at
MJD $55403.3\pm0.4$ (Corbet et al. 2011). A coincident X-ray
flux was found using \emph{Swift} X-ray telescope (XRT) observations as well as a star of magnitude B2 which in
turn coincides with the \emph{Swift/XRT} detection. We present the
results of the analysis of 5.78 Ms \emph{INTEGRAL/ISGRI} data on the
source 1FGL J1018.6-5856 (Li et al. 2011).
\begin{table}[t]
\scriptsize
\begin{center}
\label{tab1}
\caption{Reduced $\chi^2$ for fitting different models to the modulation fraction and the peak flux in X-rays.
%Radio and shifted-phase parameters fix the period in 1667 days.
}
\vspace{5pt}
\small
\begin{tabular}{lllll}
\hline
 & Constant & Linear & Radio & Shifted\\
\hline
Modulation Fraction & 88.2 / 7   & 38.0 / 6   &  42.1 / 6 & 1.1 / 5  \\
Peak Flux & 212.8 / 7   &  114.8 / 6  &  91.8 / 6 & 4.9 / 5  \\
\hline
\end{tabular}
\end{center}
\end{table}
The available \emph{INTEGRAL} observations when 1FGL J1018.6-5856 had offset angle less
than 14, adding up to a total exposure time of 5.78 Ms. Our data set covers from 2003 January 11 to 2009 November 20 . An \emph{INTEGRAL} detection of \f\ is derived by
combining all the \emph{INTEGRAL/ISGRI} data, with a significance level of 5.4$\sigma$
and an average intensity of 0.074 counts/s in the 18-40 keV
band (see Figure 2, left panel). We only obtain 1.63 $\sigma$ in
the 40-100 keV. We combine the images
from different ScWs based on orbital phase bins and produce an
orbital light curve. We show that the \emph{INTEGRAL} orbital light curve
of 1FGL J1018.6-5856 together with the \emph{Fermi/LAT} periodicity in Figure 2, right panel.
It hints a trend of having an anti-correlation between the hard X-ray emission and the \emph{Fermi/LAT} periodicity.
This is in line with the results for LS 5039 (see, e.g.,
Hoffmann et al. 2009) where the hard X-ray emission as
measured with \emph{INTEGRAL} is fully anti-correlated with
the GeV emission as measured by \emph{Fermi/LAT} (Abdo et al.
2009)--emphasizing a possible physical similarity of the two
sources. However, the scarcity of counts makes it difficult to have a definitive proof of the variability and anti-correlation: a
constant fit to the count rate yields a reduced $\chi^{2}$ of 14.09/4,
suggesting that the significance of variability is at only 2.7$\sigma$ level.

\begin{figure*}[t]
\centering
  \includegraphics[angle=0, scale=0.4] {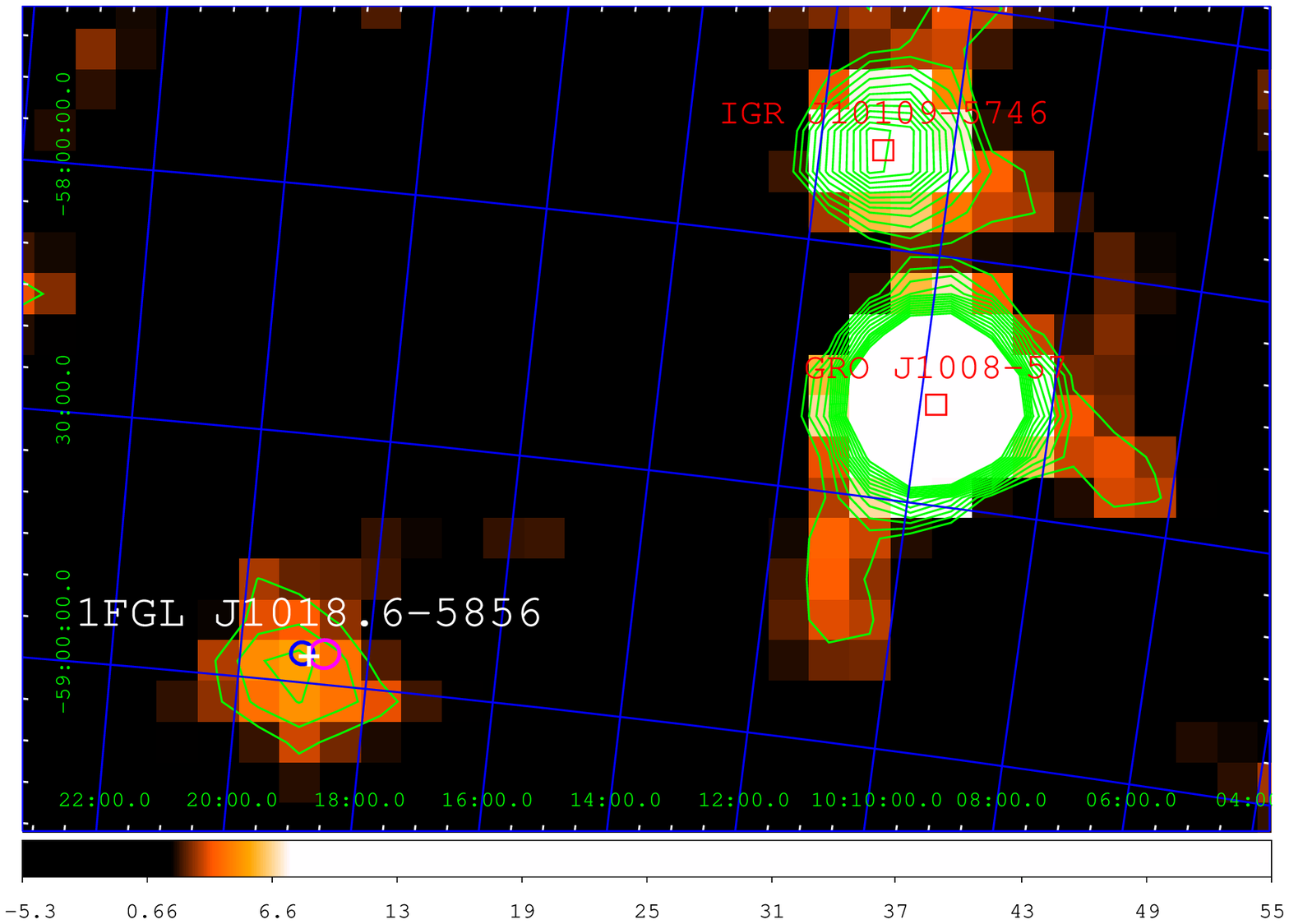}
    \includegraphics[angle=0, scale=0.36] {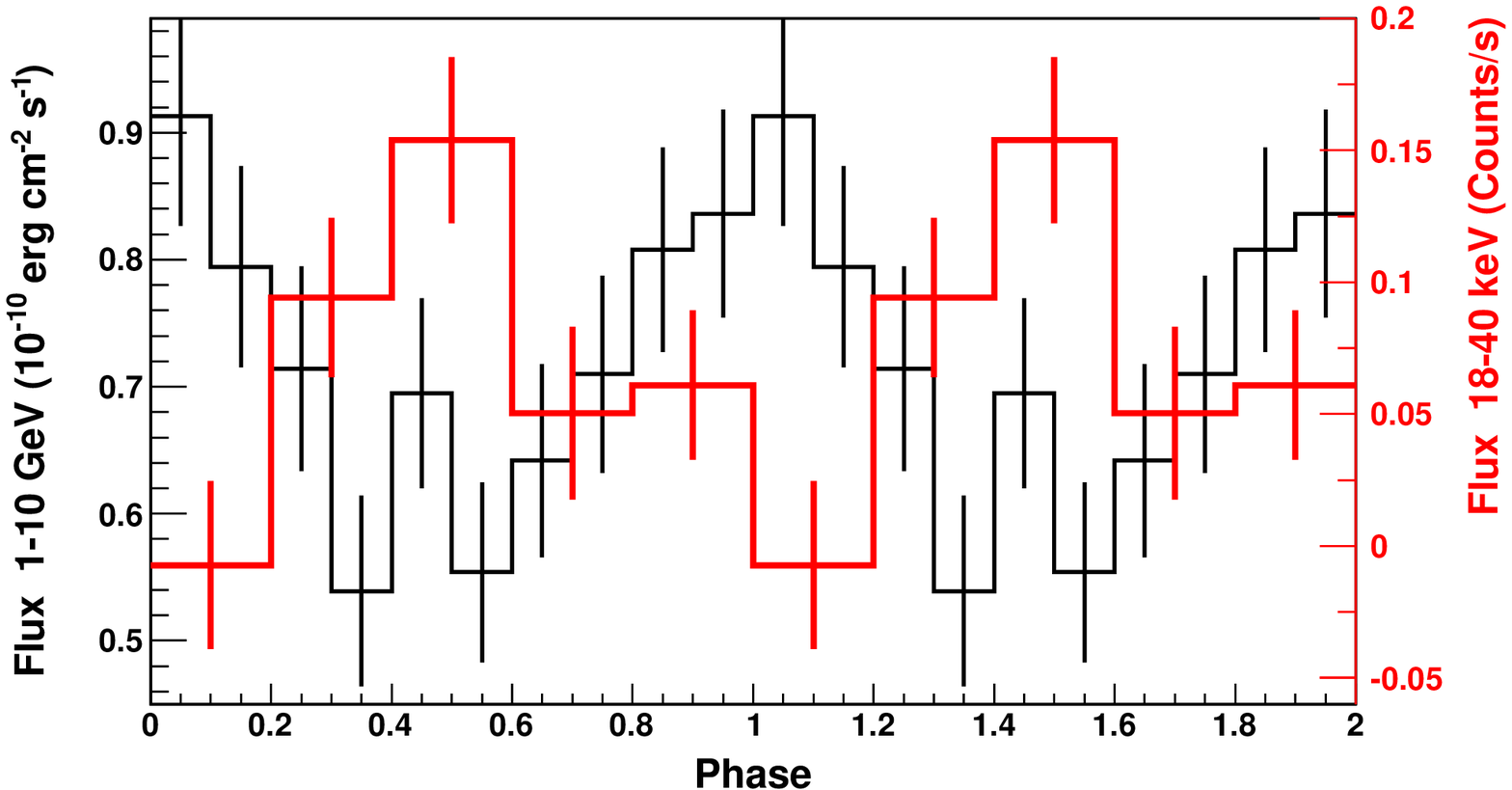}

 \caption{Left: Mosaic image of the 1FGL J1018.6-5856 sky region in the 18-40 keV band. The strongest source is GRO
J1008-57 whereas the faintest one visible only in the image is \f\. The significance level is given by the color scale, with the contours starting at 3$\sigma$, and following steps of 1$\sigma$. The position of \f\ (magenta) as well as its updated center following the 2FGL Fermi-LAT Catalog is shown, while the cross represents the counterparts identified using XRT (from Corbet et al. 2011). Right: Orbital light curve of \f\ by \emph{INTEGRAL/ISGRI} in 18-40 keV (red) and by \emph{Fermi/LAT} in 1-10 GeV (black)
}
\label{super-orb}
\end{figure*}

\section{INTEGRAL source \uu as a possible gamma-ray transient}

    In recent years, a number of unidentified, transient gamma-ray
 sources were discovered in the Galactic plane, especially by \emph{AGILE}. Many of them have also been suggested to have a possible
 binary nature (see, Table 1 in Li et al. 2012b). All
 candidates have been suggested to have a possible high mass X-ray
 binary (HMXB) counterpart. Two gamma-ray transients (AGL J1037-5708 and GRO J1036-55) are located in the same region of the
 sky and spatially associated with HMXB 4U 1036-56 (see Figure 3). \uu is identified with a B0 III--Ve star LS 1698 at $\sim$5 kpc (Motch et al. 1997) and hosting a pulsar with a period of 853.4$\pm$0.2 (La Palombara et al. 2009). We report on long-term analysis of {\it INTEGRAL} data on 4U 1036--56, examining the possibility of its association with the unidentified transient $\gamma$-ray sources AGL J1037--5708 $\&$ GRO J1036--55 (Li et al. 2012b).

    For the INTEGRAL analysis in this paper, we use all public \is and \je data for which \uu has offset angle less than 9$^o$ and 5$^o$, respectively. The data covers from 2003 January to 2009 November, adding up to a total exposure time of 4.42 Ms for \is and 1.02 Ms for \je. \uu is detected by \is with a
significance level of 11.2 $\sigma$ and an average intensity of 0.180$\pm$0.016 counts s$^{-1}$ in the 18--60 keV band.
Figure 3 shows the \is and \je mosaic image of the \uu sky region.
\uu is detected by \je at 3-20 keV at a significance of 4 $\sigma$, with an average intensity of $0.88\pm0.22 \times 10^{-4}$ counts/cm$^{2}$/s.

We investigated the \is long-term light curve of \uu on the ScW timescale in the 18-60 keV band. During most of the time, \uu is
not significantly detected by \is. However, an outburst is discovered by
\is between MJD 54142 (2007 Feb. 11) and MJD 54147 (2007 Feb. 16). This outburst at MJD 54144 is significantly detected
by \is having a significance of 30.4 $\sigma$ and an average intensity of 2.589$\pm$0.085 counts s$^{-1}$ in the 18--60 keV band,
over a 199 ks exposure ($\sim$1/22 of the total exposure on the source).
The \je detection in the outburst period is significantly made at 10.1 $\sigma$ with an average intensity of $0.194\pm0.019\times10^{-2}$
counts/{cm$^{2}$}/s in the 3--20 keV band. Out of the outburst period, \uu is detected by \is only with a significance of 5.7 $\sigma$
under a total exposure time of 4.4 Ms in the 18--60 keV band.
The average flux is 0.094$\pm$0.018 counts s$^{-1}$ in the 18--60 keV band (27 times dimmer than in outburst).
\je does not detect \uu during quiescence, yielding only 2.58 $\sigma$ in the 3-20 keV, under a total exposure of 0.99 Ms.
We extract an energy spectrum both from \je and \is and fit with absorbed blackbody, an absorbed powerlaw, and an absorbed cutoff powerlaw model. We have found that only an absorbed cutoff powerlaw model could yield an acceptable fit. According to an F-test, the probability of refusing the cutoff to a simple power-law is 4.78$\times$ $10^{-4}$, corresponding to a significance of 3.5 $\sigma$. Based on the spectrum parameters of the {\it INTEGRAL} outburst and assuming a source distance of 5 kpc (Motch et al. 1997), the luminosity derived in 2--10 keV band is ${5.16}^{+0.72}_{-0.63}\times$ $10^{35}$ erg s$^{-1}$.

The positional coincidence as well as the variability timescales make it possible to entertain the hypothesis that the high-energy transients and \uu are related. We have shown that that it is theoretically feasible that HMXBs produce gamma-ray emission during  periods of X-ray activity induced by accretion (Li et al. 2012). Hence it is in principle plausible that \uu and the {\it AGILE} flares are related.

\begin{figure*}[t]
\centering
\includegraphics[angle=0, scale=0.38] {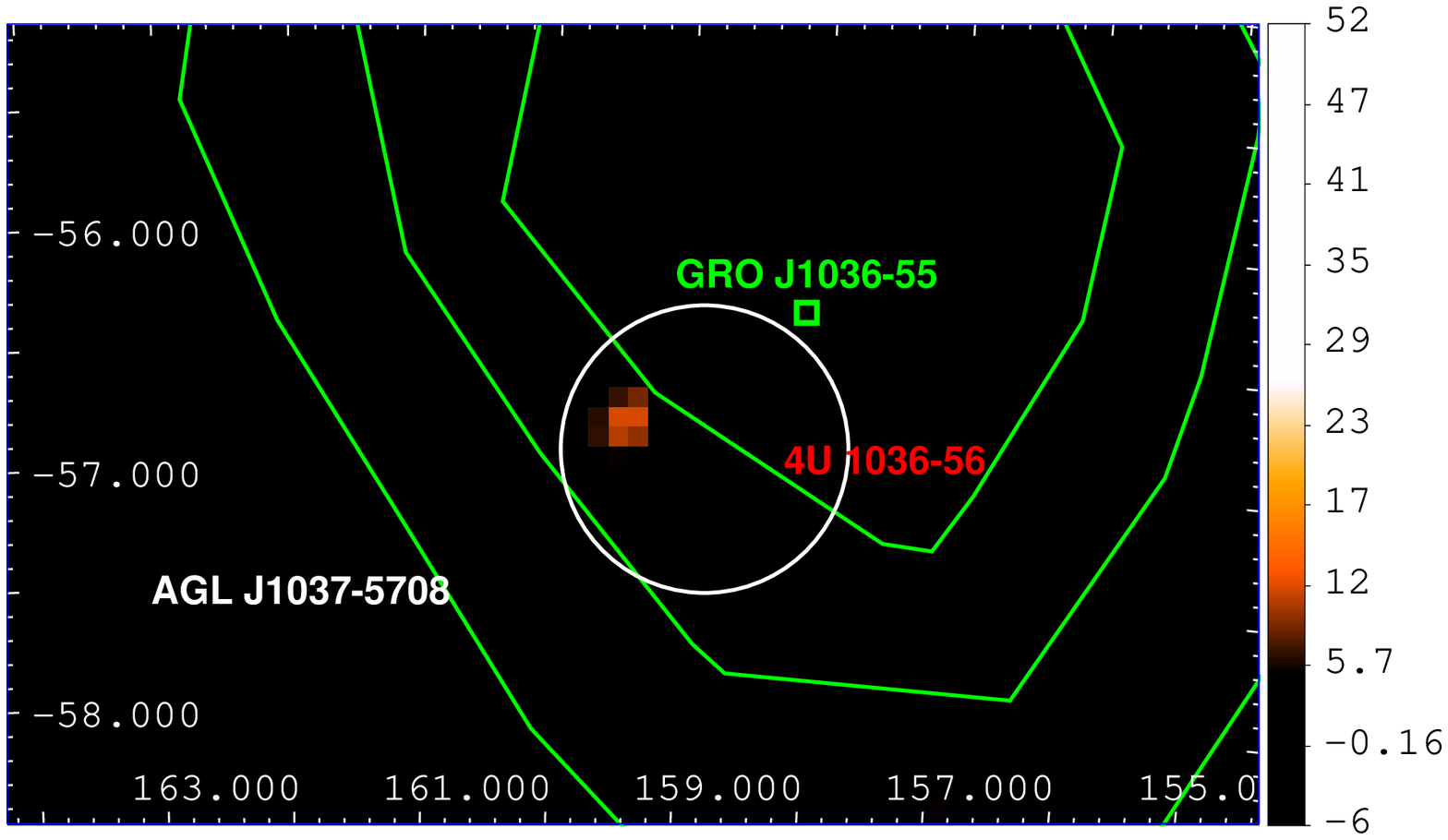}
\includegraphics[angle=0, scale=0.38] {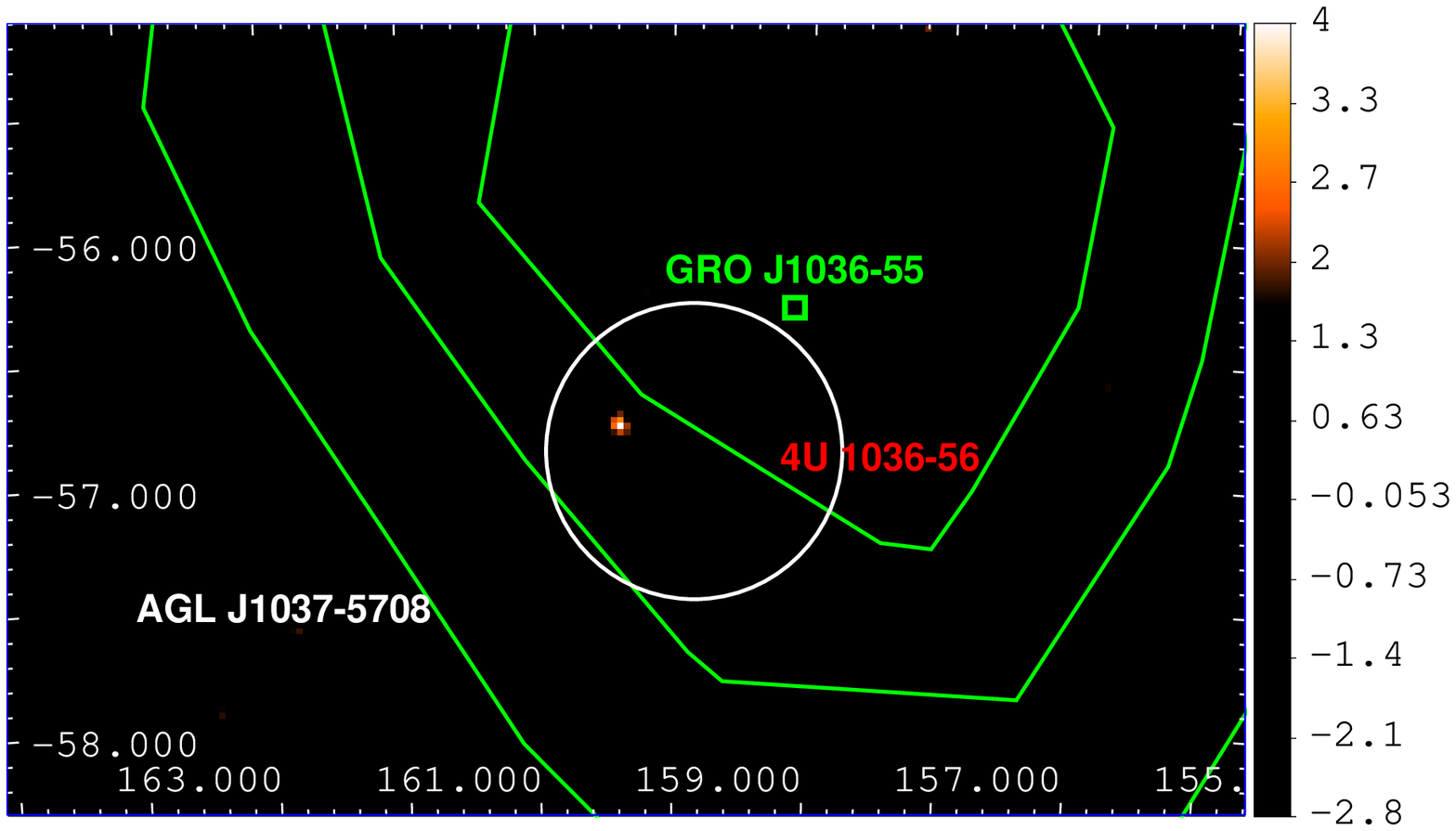}
\caption{Mosaic image of the \uu\ sky region, derived by
combining all \is data (18--60 keV, left panel) and
combining all \je data (3--20 keV, right panel). The
position of the transient {\it AGILE} source AGL J1037--5708 is plotted with its 95$\%$ error region (white). Another transient source, GRO J1036--55 is shown in sky region with its 1, 2, 3 $\sigma$ uncertainty location (green). The significance level is given by the color scale. Corresponding significance and
color can be found in the right color bar. The X-- and Y--axis are RA and Dec. in
units of degrees.}
\label{mos-1}
\end{figure*}

\section{INTEGRAL non-detection of HESS J0632+057 and AGL J2241+4454 }

\h\ is a gamma-ray binary with 321 days orbital periods (Bongiorno et al. 2011) and a companion of B0 Vpe star (Casares et al. 2011). It is detected in radio (Skilton et al. 2009), soft X-ray (Falcone
et al. 2010) and TeV (Aharonian et al. 2007). We look for its hard X-ray counterpart with \emph{INTEGRAL}. At an offset angle of 14 degree,
\h\ is observed by \is with a exposure of 0.81 Ms and by \je with 15 ks exposure from 2003 to 2011. But \h\ is not detected in the 18--60 keV energy range by \is nor in 3-35 keV range by \je.
\agl is a transient gamma-ray source detected by the \emph{AGILE} satellite above 100 MeV (Lucarelli et al. 2010). It is possibly associated with Be star MWC 656 with an orbital period of $60.37\pm0.04$ days (Williams et al. 2010). It is a gamma-ray binary candidates (Li et al. 2012b). We try to identified its counterpart in hard X-ray by \emph{INTEGRAL}. The available \emph{INTEGRAL/ISGRI} data on \agl\ with an offset angle of 14 degree leads to a total exposure of 2.1Ms.
The detection significance is only 1.37 sigma in 18-60 keV. As a result, we do not have a detection of \agl. \\

{\small This work is supported by 973 program 2009CB824800 and the National Natural
Science Foundation of China via NSFC-11233003, 11103020, 11133002, 11073021 and
11173023. Research done in the framework of the grants AYA2012-39303, AYA2009-07391,
SGR2009-811, and iLINK2011-0303. We acknowledge Nanda Rea, Daniela Hadasch, Andrea Caliandro, Alessandro Papitto, and Yupeng Chen for joint work on these sources.


\begin{thebibliography}{99}
\bibitem{abdo2009}Abdo, A. A., Ackermann, M., Ajello, M., et al., 2009, \emph{ApJ}, 706, L56
\bibitem{abdo2009}Aharonian, F. A. et al., 2007, \emph{A$\&$A}, 469, L1
\bibitem{abdo2009}Bongiorno, S. D., Falcone, A. D., Stroh, M., et al., 2011, \emph{ApJ}, 737, L11
\bibitem{abdo2009}Casares, J., Rib$\acute{o}$, M., Ribas, I., et al., 2012, \emph{MNRA}S, 421, 1103
\bibitem{abdo2009}Chernyakova, M., Neronov, A., Molkov, S., et al., 2012, \emph{ApJ}, 747, 29
\bibitem{abdo2009}Corbet, R. H. D., Cheung, C. C., Kerr, M., et al., 2011, \emph{ATel}, 3221
\bibitem{abdo2009}Falcone, A. D., Grube, J., Hinto,n J.,et al., 2010, \emph{ApJ}, 708, L52
\bibitem{abdo2009}Gregory, P. C., 2002,\textit{ApJ}, 575, 427
\bibitem{abdo2009}Hadasch, D., Torres, D. F., Tanaka, T., et al., 2012, \textit{ApJ}, 749, 54
\bibitem{abdo2009}Hoffmann, A. D., Klochkov, D., Santangelo, A., et al., 2009, \emph{A$\&$A}, 494, L37
\bibitem{abdo2009}La Palombara, N., Sidoli, L., Esposito, P., Tiengo, A., $\&$ Mereghetti, S., 2009,\emph{ApJ}, 505, 947
\bibitem{abdo2009}Li, J., Torres, D. F., Zhang, S., et al., 2012a,\textit{ApJ}, 744, 13
\bibitem{abdo2009}Li, J., Torres, D. F., Zhang, S., et al., 2012b,\textit{ApJ}, 761, 49
\bibitem{abdo2009}Li, J., Torres, D. F., Zhang, S., et al., 2011,\textit{ApJ}, 733, 89
\bibitem{abdo2009}Lucarelli F. et al., 2010, \emph{ATel}, 2761
\bibitem{abdo2009}Motch, C., Haberl, F., Dennerl, K., Pakull, M.,$\&$Janot-Pacheco, E., 1997, \emph{A$\&$A}, 323, 853
\bibitem{abdo2009}Skilton, J. L. et al., 2009, \emph{MNRAS}, 399, 317
\bibitem{abdo2009}Torres, D. F., et al., 2012,\textit{ApJ}, 744, 106
\bibitem{abdo2009}Papitto, A., Torres, D. F. $\&$ Rea, N., 2012, \textit{ApJ}, 756, 188
\bibitem{abdo2009}Williams, S. J., Gies D. R., Matson, R. A., et al., 2010, \emph{ApJ}, 723, L93





\end{thebibliography}
\end{document}